\documentclass{article}
\usepackage{arxiv}
\usepackage[utf8]{inputenc} 
\usepackage[T1]{fontenc}    
\usepackage{hyperref}       
\usepackage{url}            
\usepackage{booktabs}       
\usepackage{amsfonts}       
\usepackage{nicefrac}       
\usepackage{microtype}      
\usepackage{graphicx}
\usepackage{doi}
\usepackage{siunitx}

\usepackage{amsmath}
\usepackage{caption}
\usepackage{subcaption}

\usepackage{amssymb}
\usepackage{xurl}
\usepackage{anyfontsize}
\usepackage{graphicx}%
\usepackage{amsmath,amssymb,amsfonts}%
\usepackage{amsthm}%
\usepackage[title]{appendix}%
\usepackage[table,xcdraw]{xcolor}%
\usepackage{textcomp}%
\usepackage{algorithm}%
\usepackage{algorithmicx}%
\usepackage{algpseudocode}%
\usepackage[numbers]{natbib}
\usepackage{tabularx}
\usepackage{subcaption}
\usepackage{siunitx}
\usepackage{changepage}
\usepackage[utf8]{inputenc} 
\usepackage[T1]{fontenc}

\graphicspath{ {./figs} }

\DeclareMathOperator{\dotprod}{\boldsymbol{\cdot}}  
\DeclareMathOperator{\nablavec}{\boldsymbol{\nabla}}  

\title{Topology-Informed Machine Learning for Efficient Prediction of Solid Oxide Fuel Cell Electrode Polarization}

\author{ Maksym Szemer \\
	AGH University of Krakow,
	Krakow, Poland \\
  \texttt{szemermaksym@student.agh.edu.pl} \\
	\And
	\href{https://orcid.org/0000-0002-2959-3044}{\includegraphics[scale=0.06]{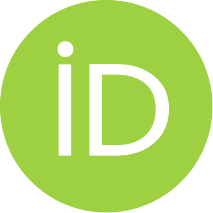}\hspace{1mm}Szymon Buchaniec} \\
	AGH University of Krakow,
	Krakow, Poland \\
	\texttt{buchaniec@agh.edu.pl} \\
	 \And
	 \href{https://orcid.org/0000-0003-1898-571X}{\includegraphics[scale=0.06]{orcid.pdf}\hspace{1mm}Tomasz Prokop}\\
     AGH University of Krakow,
	Krakow, Poland \\
	\texttt{prokopt@agh.edu.pl} \\
	 \And
	\href{https://orcid.org/0000-0003-4911-5880}{\includegraphics[scale=0.06]{orcid.pdf}\hspace{1mm}Grzegorz Brus \thanks{Corresponding author: Grzegorz Brus {brus@agh.edu.pl}}}\\
	AGH University of Krakow,
	Krakow, Poland \\
	 \texttt{brus@agh.edu.pl}
}

\begin{document}
\maketitle
\begin{abstract}
Machine learning has emerged as a potent computational tool for expediting research and development in solid oxide fuel cell electrodes. The effective application of machine learning for performance prediction requires transforming electrode microstructure into a format compatible with artificial neural networks. Input data may range from a comprehensive digital material representation of the electrode to a selected set of microstructural parameters. The chosen representation significantly influences the performance and results of the network. Here, we show a novel approach utilizing persistence representation derived from computational topology. Using 500 microstructures and current-voltage characteristics obtained with 3D first-principles simulations, we have prepared an artificial neural network model that can replicate current-voltage characteristics of unseen microstructures based on their persistent image representation. The artificial neural network can accurately predict the polarization curve of solid oxide fuel cell electrodes. The presented method incorporates complex microstructural information from the digital material representation while requiring substantially less computational resources (preprocessing and prediction time $\approx \SI{1}{min}$) compared to our high-fidelity simulations (simulation time $\approx \SI{1}{\hour}$) to obtain a single current-potential characteristic for one microstructure.
\end{abstract}

\section{Introduction}
A Solid Oxide Fuel Cell's (SOFC) capacity to convert chemical energy into electricity depends on the complex microstructure of its electrodes. This intricate catalyst structure enables the essential reactions that power the cell, optimizing the ion, electron, and gas transfer pathways while minimizing the resistance to these flows. Understanding the relationship between microstructural features and engineering them is crucial for enhancing the efficiency of the entire energy system. Various research groups focus on these electrodes' material science and chemical engineering to unveil new strategies to optimize the microstructural elements, thereby contributing to more robust and energy-efficient fuel cell technologies.

In the early paper by Bozorgmehri and Hamedi \cite{Bozorgmehri.20124u7}, an artificial neural network (ANN) and a genetic algorithm (GA) are employed to model and optimize the parameters of intermediate-temperature solid oxide fuel cells. The parameters optimized in the study are anode support layer thickness and its porosity, electrolyte thickness, and cathode functional layer thickness. The ANN model, trained using experimental data, not only predicts the performance of the SOFC without relying on physical models but also presented the potential to revolutionize SOFC research. The GA is used to determine the optimal values for four key cell parameters, resulting in the highest maximum power density under various constraints. In the paper by Ahmed M. Nassef et al. \cite{Nassef.2019rp}, an optimization algorithm called radial movement optimizer is used to maximize the performance of solid oxide fuel cells. The study models the SOFC using Artificial Neural Networks based on experimental data and identifies optimal operating parameters such as anode support layer thickness, anode porosity, electrolyte thickness, and cathode interlayer thickness. The proposed optimization method significantly enhances the cell’s maximum power density, achieving a 33.8\% and 17.28\% increase over experimental and genetic algorithm results, respectively. To this moment, the inclusion of microstructure in optimization was limited to porosity. This change was made in the research by  Buchaniec et al. \cite{Buchaniec.2019}, who use evolutionary algorithms to optimize the microstructure of a solid oxide fuel cell anode. The study focuses on 16 microstructural parameters such as phase volume fraction, connectivity, the tortuosity factor, triple phase boundary (TPB) length density, mean particle diameter, and its standard deviation to enhance the electrochemical performance of the anode, which is typically a composite of nickel and yttria-stabilized zirconia (Ni-YSZ). The optimized parameters for the SOFC anode microstructure include the anode composition (volume fractions of Ni and YSZ) and the mean pore particle diameter. The results demonstrate that the optimized microstructure significantly improves cell performance compared to conventional designs.

A step forward from optimizing a set of microstructural parameters is direct morphology optimization. An example of such research for porous mixed conducting cathodes for solid oxide fuel cells was studied by He et al. \cite{He.2020,He.2021}. The optimized La$_{0.6}$Sr$_{0.4}$Co$_{0.2}$Fe$_{0.8}$O$_3$ cathode microstructure showed a \SI{40}{\percent} increase in total reaction current compared to the original structure. The study also identified universal optimal geometric features, such as small particles and a monotonically decreasing solid volume fraction from the electrolyte along the thickness direction. Later similar optimization was conducted for the anode by the same group \cite{He.2022}. Another morphology optimization of cathode microstructure was conducted by Qiangqiang Li et al. \cite{Li.2020}, to address the issue of thermal mismatch in solid oxide fuel cells. The authors used the asymptotic homogenization method and three-phase material topology optimization. They successfully designed two orthogonal microstructures that minimize thermal mismatch and align the thermal expansion coefficients of the cathode with those of the electrolyte layer, thereby reducing stress and potential fractures. 

A study by Timurkutluk et al. \cite{Timurkutluk.2023bbw} used an artificial neural network to model active triple-phase boundaries in solid oxide fuel cell electrodes based on particle size. Synthetic SOFC electrode microstructures were generated using Dream.3D software, and active TPB densities were measured using COMSOL software for ANN training and validation. Among the three learning methods tested, Bayesian regulation (BR) with one hidden layer and five neurons performed the best, showing a mean relative error of 0.036. The ANN model accurately predicted active TPB densities for new particle sizes, demonstrating its potential as a predictive tool for SOFC electrode design, which can be further improved by considering additional microstructural properties. In the article by Xu et al. \cite{Xu.2022}, the microstructure was partially included to improve the performance of solid oxide fuel cells. In this research, the authors combined three-dimensional computational fluid dynamic modeling, artificial neural networks, and genetic algorithms. The study highlights that the ANN achieved high accuracy in predicting SOFC performance, facilitating power density optimization. The microstructure of SOFCs, including parameters like particle size and porosity, plays a crucial role in enhancing current density and overall performance. Recently, Gnatowski et al. \cite{GNATOWSKI202311823} included the microstructure of electrodes using a hybrid model. The neural network, trained on experimental data, dynamically updates charge transfer coefficients based on temperature and current, improving the accuracy of electrochemical reaction models. It is important to note that the network was used to tune physical parameters in the model, not to accelerate the computation required for simulations used in evolutionary optimization.

Machine learning has proven to be a valuable computational tool in accelerating research and development for solid oxide fuel cell electrodes. However, as highlighted in a recent review, performance prediction using neural networks that consider microstructure remains an ongoing challenge \cite{Brus.2024.review}. This difficulty may stem from the need to convert electrode microstructures into a format compatible with artificial neural networks. Typically, this is achieved through microstructural parameters that simplify the complexity of the microstructure. The choice of these parameters significantly impacts the network's performance and outcomes. A potential solution to this issue is the use of persistent representations derived from computational topology, which incorporates complex microstructural information available from digital material representations, suitable for machine learning applications\cite{Rammal.2022}. 

The concept of persistence homology (PH) was first publicly presented during the 41st Annual Symposium on Foundations of Computer Science in November 2000 and published in 2002 by Edelsbrunner et al., who studied the topological persistence of the incrementing simplicial complex sequence in $\mathbb{R}^3$ \cite{Edelsbrunner}. Then Zomorodian and Carlsson presented the mathematical foundations of persistent homology in 2004\cite{Zomorodian_Carlsson_2004}.

Apart from material research, in the recent time, topological data analysis has been successfully introduced into fields such as feature selection, cosmology\cite{Wilding_Nevenzeel_van_de_Weygaert_Vegter_Pranav_Jones_Efstathiou_Feldbrugge_2021, Heydenreich_Bruck_Harnois-Deraps_2021}, protein analysis\cite{Gameiro_Hiraoka_Izumi_Kramar_Mischaikow_Nanda_2015, Xia_Wei_2014, Xia_Feng_Tong_Wei_2015} and medicine\cite{Singh_Farrelly_Hathaway_Leiner_Jagtap_Carlsson_Erickson_2023}. Topological data analysis has especially resulted in great progress in the last one mentioned.

Over the last few years, the rapid development of deep learning and artificial intelligence, in general, has had an impact on medicine and has made an influence to increase the accuracy and speed of diagnoses at an early stage of many diseases or conditions.
Techniques such as computational tomography, MRI, or ultrasonography are commonly used in diagnosing diseases, however, their uses involve introducing multidimensionality or metric mismatches related to hardware calibration, which may cause problems with deep learning or other machine learning algorithms\cite{Nielson_Cooper_Yue_Sorani_Inoue_Yuh_Mukherjee_Petrossian_Paquette_Lum_2017}. In medicine, topological data analysis is used inter alia neurology, cardiology, hepatology and oncology\cite{Han_Zhou_Fei_Sun_Wang_Chen_Chen_Wang_Tang_Ge._2020, Malek_Alias_Razak_Noorani_Mahmud_Zulkepli_2023}. 

Detection of microcalcifications during mammography is crucial to detecting early-stage breast cancer. The feature characterized by PH, noise filtering, and machine learning models, allowed to significantly increase in the early detection of breast cancer in women. Additionally, it was found that machine learning techniques based on PH are characterized by greater accuracy in classification\cite{Malek_Alias_Razak_Noorani_Mahmud_Zulkepli_2023}, moreover, these techniques give surprisingly positive results in machine learning based on limited training sets\cite{Kofler_Dewey_Schaeffter_Wald_Kolbitsch_2020}.

There is a wide range of publications related to PH applied in material science\cite{Nakamura_Hiraoka_Hirata_Escolar_Nishiura_2015, Obayashi_Nakamura_Hiraoka_2022, PAWLOWSKI2023100256, structural_hirata_2020, percolating_robins_2016, persistent_suzuki_2023, improving_prokop_2023, design_ge_2012, threedimensional_prokop_2018}, however, they mostly use PH as a tool to extract topological features, e.g. to show changes occurring in the examined object\cite{PAWLOWSKI2023100256}, without linking them to a quantitative feature. One of the few works that deals with the above-mentioned approach is Shuseki et al. article\cite{Shuseki2024} in which they explain the differences in the glass-forming ability through the topological similarity of the studied structures.

A literature survey can be summarized as follows: The topological optimization of microstructure features in solid oxide fuel cell electrodes is a relatively new and burgeoning field. This optimization aims to meet specific criteria, primarily associated with the electrochemical performance of the cells. While neural networks could potentially assist in this task, there are currently no neural networks that incorporate complete or comprehensive information about microstructures. This study addresses this gap by demonstrating the feasibility of using computational topology, specifically persistent homology, in an artificial neural network to integrate complete microstructural feature information. This enables the network to accurately predict the performance of an electrode.

\section{Methods}

\subsection{Theory}\label{sec:theory}
To present the process of creating our chosen topological descriptors, persistent images, it is necessary to introduce several concepts related to topological data analysis. 

Algebraic topology captures characteristics of metric spaces, but in certain situations, we are not interested in the precise geometry of these spaces, instead, we want to understand basic properties, such as the number of components or existence of voids\cite{Berry_Chen_Cisewski-Kehe_Fasy_2020}.
To do that we can use homology, which is an algebraic topological invariant, that represents $k$-dimensional quantities associated with topological space that do not change under continuous deformations of the space X, which are encoded in an $k$-th homology group of X, noted as $H_k(X)$\cite{ Adams_Chepushtanova_Emerson_Hanson_Kirby_Motta_Neville_Peterson_Shipman_Ziegelmeier_2016, Simon.2021}. 
The rank of $H_k(X)$ counts the number of k-dimensional holes in $X$, where, for example, for $k =0,1,2$ it corresponds to connected components, rings and cavities respectively. The rank of the homology group is referred to as the Betti number which is the rank of that homology group:
\begin{equation}\label{eq:betti_number}
    \beta_k = \mathrm{dim} \space H_k(X)
\end{equation}

Equivalent topological spaces have the same $\beta_k$ for corresponding $k$, hence Betti numbers are used to distinguish topological spaces. In this paper, we only consider $k = 0,1,2$, but Betti numbers can be consequently applied for higher dimensional structures.

\begin{figure}[ht!]
\begin{subfigure}{\textwidth}
    \centering
    \includegraphics{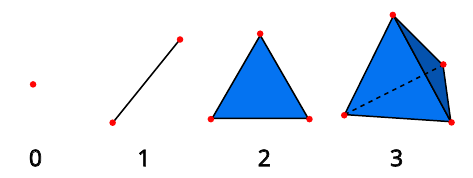}
    \caption{\textbf{Graphical representation of few simplices.} 0-simplex, 1-simplex, 2-simplex, and 3-simplex, which are respectively, vertex, edge, triangle and tetrahedron}
    \label{fig:first_few_simplices}
\end{subfigure}
\vspace{1em}
\begin{subfigure}{\textwidth}
    \centering
    \includegraphics{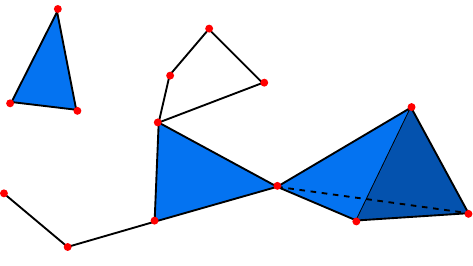}
    \caption{\textbf{A simplicial 3-complex.} Betti numbers of that given complex are as follows: $\beta_0 = 2$, $\beta_1 = 1$, $\beta_2 = 1$, $\beta_i = 0 \quad \forall i \geq 3$}
    \label{fig:simplicial_complex_example}
\end{subfigure}
        
\label{fig:figures}
\caption{\textbf{Graphical representation of first four simplices and a simplicial 3-complex with Betti numbers}}
\end{figure}

Let's introduce two concepts, simplices and simplicial complexes\cite{Dey_Wang_2022}.
Simplices are generalizations of the notion of a triangle to all dimensions. The first few simplices are shown in \autoref{fig:first_few_simplices}.

Formally for $m\geq0$, a $m$-simplex $\sigma$ in a Euclidean space $\mathbb{R}^n$ is the convex hull of set $P$ of $m + 1$ affinely independent points in $\mathbb{R}^n$.

A simplicial complex $K$, is a set of simplices that satisfies two restrictions:
$K$ contains every face of each simplex in $K$ and for any two simplices $\sigma, \tau \in K$, their intersection $\sigma \cap \tau$ is either empty or face of both $\sigma$ and $\tau$

The notion of 'persistence' refers to the appearance and disappearance of the observed properties during the so-called filtration process, which is used to count Betti numbers of a given topological space. The process of filtration is slightly different based on which type of data is used, when working with topological data analysis in material science, two types of data, point cloud and cubical could be most often encountered\cite{Obayashi_Hiraoka_Kimura_2018}.

A point cloud is commonly used to express atomic configurations of materials and is defined as $P = {x_i \in \mathbb{R}^n}$ in Euclidean space $\mathbb{R}^n$. To perform the process of filtration it is convenient to define the r-ball model\cite{Obayashi_Hiraoka_Kimura_2018}:
\begin{equation}\label{balls}
  P_r = \bigcup_{i=1}^m B_r\left(x_i\right)  
\end{equation}
where $B_r\left(x_i\right) = \{ y \in \mathbb{R}^n : \|y - x_i\| \leq r \}$ is euclidean ball centered at data points $x_i$ with radius $r$. Using this construction, when radius $r$ is small, $P_r$ has the same topology as $m$ separated points, increasing the value of $r$ proceeds appearing and disappearing of holes, and when $r$ is big, $P_r$ has respectively the same topology as one point.

The simplicial complex that encodes the topological properties of $P_r$ is the Čech complex $\check{C}ech\left(P, r\right)$\cite{Kim_Shin_Chazal_Rinaldo_Wasserman_2020}.
If $\sigma = \{x_0, ..., x_k \}$ is simplex created from points belonging to the set P, Čech complexes defined as:
\begin{equation}\label{eq:cech_complex}
\check{C}ech\left(P, r\right) = \Bigl\{ \sigma \in P :\bigcap_{s=0}^k B_r\left(x_s\right) \neq \varnothing \Bigr\}
\end{equation}

As computing the Čech complex involves computing all possible intersections of the balls, to speed up the computation, we can only check the pairwise distances between the data points and thus build the Vietoris-Rips complex $Rips\left(P, r\right)$:
\begin{equation}\label{eq:rips_complex}
    Rips\left(P, r\right) = \Bigl\{ \sigma \in P : B_r\left(x_s\right) \cap B_r\left(x_t\right) \neq \varnothing \Bigr\}
\end{equation}
for all $x_s, x_t \in \sigma$.

 Filtration $\mathcal{F} = \mathcal{F}(K)$ of a simplicial complex $K$ is a nested sequence of its subcomplexes\cite{Dey_Wang_2022}:
\begin{equation}\label{eq:filtration}
    \mathcal{F}: \varnothing = K_0 \subseteq K_1 \subseteq ... \subseteq K_{n-1} \subseteq K_n = K
\end{equation}
which could also be defined using the $r$-ball model:
\begin{equation}\label{eq:ball_filtration}
    \mathcal{F} = \{K_r: r \in \mathbb{R}\}
\end{equation}
where $K_r$ is Čech or Vietoris-Rips complex.
For more details about filtration we refer to Nielsen\cite{Nielsen_Barbaresco_2019} and Dey\cite{Dey_Wang_2022} books, while a more engineering perspective is provided by Obayashi\cite{Obayashi_Hiraoka_Kimura_2018} and Pawłowski\cite{PAWLOWSKI2023100256}.

The cubical set is a typical way to mathematically express digital images which was previously used for analysing properties of microstructures\cite{Dlotko_Wanner_2016}. This approach was used in this paper.
The cubical set is not limited only to 2-dimensional(pixel) or 3-dimensional(voxel) data. The general definition was provided by Kaczynski\cite{Kaczynski_Mischaikow_Mrozek_2010}.
Let $I \subset \mathbb{R}$ be elementary interval that is:
    \begin{equation}\label{eq:elementary_interval}
        I = \left[l, l + 1\right] \quad \text{or} \quad I = \left[l, l\right]
    \end{equation}
    for some $l \in \mathbb{Z}$. An elementary cube $Q$ is defined as $Q = I_1 \times \dots \times I_n \subset \mathbb{R}^n$, where $I_i$ is an elementary interval. Subset $X \in \mathbb{R}^n$ is cubical if $X$ can be expressed as the sum of elementary cubes in $\mathbb{R}^n$.
With a function $f: \mathcal{K}^n \rightarrow \mathbb{R}$, we can build a nested sequence of cubical sets:
\begin{equation}\label{eq:cubical_filtration}
    X_t = \bigcup \{Q \in \mathcal{K}^n : f(Q) \leq t\}
\end{equation}
where $\mathcal{K}^n$ is the set of elementary cubes in $\mathbb{R}^n$. In application, Manhattan distance or grey-scale is often used as function $f$. Such defined cubical sets also lead to filtration $\mathcal{F} = \{ X_t : t \in \mathbb{R} \}$\cite{Obayashi_Hiraoka_Kimura_2018}.

For clarity, we will refer to the values $r$ (\autoref{eq:ball_filtration}) and $t$ (\autoref{eq:cubical_filtration}) that appears in filtration definitions as generators.
The value of the generator of appearance and disappearance is called respectively birth $b$ and death $d$ time, thus the $d-b$ is the persistence of the observed feature and each pair could appear multiple times. Each birth-death pair expresses that a k-dimensional hole appears in X, and then disappears.
Given that a decomposition Persistence Diagram and Persistence Barcode can be produced, in the process of creating a Persistence Image, Persistence Diagrams (PD) are needed, so in this work, we focused on that approach.
Such a decomposition is unambiguous.
The Persistence Diagram is a multiset of birth-death pairs generated during filtration, given that, $k$-th Persistence Diagram $D_k(X)$ of $X$ is defined as multiset:
\begin{equation}\label{eq:persistence_diagram}
    D_k\left(X\right) = \{\left(b_i, d_i\right) \in \Delta : i = 1, ..., p \}
\end{equation}
where $\Delta = \{ \left(b,d\right) \in \mathbb{R}^2 : b<d \}$.

In most approaches, pairs with long persistence, are key properties, while those with very short persistence are considered as noise\cite{Cohen-Steiner_Edelsbrunner_Harer_2006}. There are some approaches where mid-value persistence properties are key\cite{Bendich_Marron_Miller_Pieloch_Skwerer_2016}.

\begin{figure}[ht]
    \centering
    \includegraphics{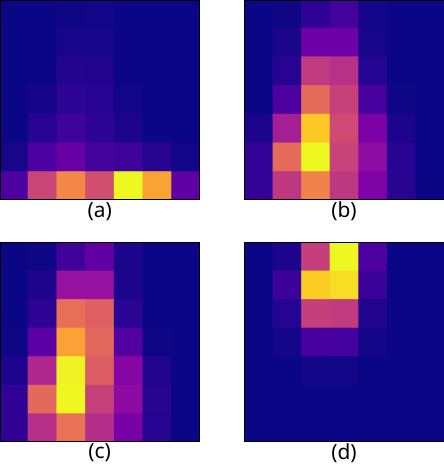}
    \caption{Comparison of PI results for different weight functions}
    \label{fig:comparision}
\end{figure}

After all those preparations, we can finally proceed to the last stage of creating persistent images.
Let us denote PD, in birth-death coordinates, as $D$. Let $T: \mathbb{R}^2 \rightarrow \mathbb{R}^2$ be a linear transformation such that $T(x, y) = (x, y-x)$, then $T(D)$ is the transformation from birth-death coordinates to birth-persistence coordinates. Next persistence surface is created by applying transformation $\rho_D: \mathbb{R}^2 \rightarrow \mathbb{R}$:
\begin{equation}\label{eq:persistence_surface}
    \rho_D\left(u\right) = \sum_{u \in T\left(D\right)} w\left(u\right)\phi_\mu\left(u\right)
\end{equation}
where $w$ is weight function, $\phi_u$ is differentiable probability density function with mean $\mu = \left(\mu_x,\mu_y\right) \in \mathbb{R}^2$. The choice of the probability distribution and weighting function affects the PI created\cite{Kusano_Fukumizu_Hiraoka_2017} which was shown on \autoref{fig:comparision}.
A commonly used probability distribution is normal distribution $g_u$, we use it in this work. The selection of the weight function $w$ is crucial to ensure the stability of the persistence surface and persistence image\cite{Edelsbrunner_2012, Adams_Chepushtanova_Emerson_Hanson_Kirby_Motta_Neville_Peterson_Shipman_Ziegelmeier_2016}. In our work, we used the weight function proposed by Kusano\cite{Kusano_Fukumizu_Hiraoka_2017}:
\begin{equation}\label{eq:weight_function}
    w\left(u\right) = \arctan\left(C \cdot \left(u_y\right)^p\right)
\end{equation}
where $C > 0, \quad p \in \mathbb{Z}_{>0}$ are parameters of weight functions, which will be treated as hyperparameters and will be selected in the cross-validation process.
Therefore, the final transformation $\rho_D$ in our case will have the form:
\begin{equation}\label{eq:surface_full}
    \rho_D\left(u\right) = \sum_{u \in T\left(D\right)} w\left(u\right) \cdot e^{\left(-\frac{\left(u_x - \mu_x\right)^2 + \left(u_y - \mu_y\right)^2}{2\sigma^2}\right)}
\end{equation}

where $u_x, u_y$ are respectively birth and persistent values. Final transformation provides to reduce the persistent surface to a finitely-dimensional vector through discretizing process.
Persistent image(PI) was defined by Adams\cite{Adams_Chepushtanova_Emerson_Hanson_Kirby_Motta_Neville_Peterson_Shipman_Ziegelmeier_2016} as $PI\left(\rho_D\right)_p := \int\int_p \rho_D \mathrm{d}y\mathrm{d}x$, however Kusano\cite{Kusano_Fukumizu_Hiraoka_2017} provides a more convenient definition as $m \times m$ matrix, where $(i, j)$-th element is integral of $\rho_D$ over pixels $P_{i,j} := \left(a_{i-1}, a_i\right] \times \left(a_{j-1}, a_j\right]$, so:
\begin{equation}\label{eq:persitence_image}
    PI\left(D\right)_{i,j} := \int_{P_{i,j}} \rho_{D}\left(u\right) \mathrm{d}u
\end{equation}

lately, in our work, we use a vectorized form of PI defined as a $m^2$-dimensional vector:
\begin{equation}\label{eq:vectorized_image}
    PIV\left(D\right)_{i+m\cdot \left(j-1\right)} := PI\left(D\right)_{i,j}
\end{equation}

in our work, this form will be referred to as a Persistent Image.
Creating PI as a topological descriptor also involves selecting mesh sizes/resolutions, and the number of pixels, however, research shows that even Persistent Images with low resolution retain topological features very well, which leads to similar performance to those with high resolution. This property is highly desirable due to the significant reduction in computational costs with a small reduction in performance. This significant property was also confirmed in our work, as shown in the next section.

\begin{figure}[htbp]
    \begin{adjustwidth}{-2cm}{-2cm}
    \centering
    \includegraphics{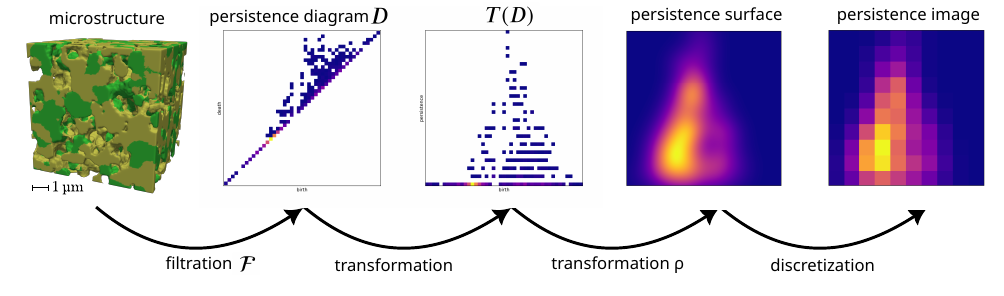}
    \label{fig:pipeline}
    \end{adjustwidth}
    \caption{\textbf{The pipeline of creating Persistence Image.}The pipeline for creating PI from source data consists of several consecutive transformations. Filtration $\mathcal{F}$ generates birth-death pairs that constitute PD. The linear transformation $T$ is responsible for the transition to birth-persistence coordinates. Then with \autoref{eq:persistence_surface} Persistence Surface is created and then discretized to the topological descriptor - Persistence Image}
\end{figure}

\subsection{Description of dataset}\label{sec:dataset}
In this section we describe our approach, starting from the describe of generating microstructures, data preprocessing and description of artificial neural network architecture.

The dataset used in this research consists of $538$ artificially generated microstructures, which were split into training, validation and test sets in ratios $0.65, 0.15, 0.2$ except for limited training dataset performance comparison in which the training set was reduced in size.
In the simulation, overpotential (voltage loss due to thermodynamic irreversibilities) for each microstructure was calculated for $8$ values of current density: $0, 500, 1000, 1500, 2000, 2500, 3000$ and $3500$, which including value for current density gives in result $3200$ training records. The mathematical and numerical model used for computing the overpotential is presented in Appendix \ref{app:micromodel}.

Each microstructure was decomposed into three cubical sets, each corresponding to one microstructure's phase, and on each of them filtration process was performed using \href{https://homcloud.dev/index.en.html}{HomCloud}, software for persistent homology which was already used in various researches\cite{Obayashi_Nakamura_Hiraoka_2022, Shimizu_Kurokawa_Arai_Washizu_2021}. We used $0$, $1$ and $2$nd Persistence Diagrams for our purpose, given that $9$ Persistence Diagrams were generated for each microstructure.

\subsection{Machine learning details}\label{sec:ml_details}
\begin{figure*}
    \centering
    \includegraphics{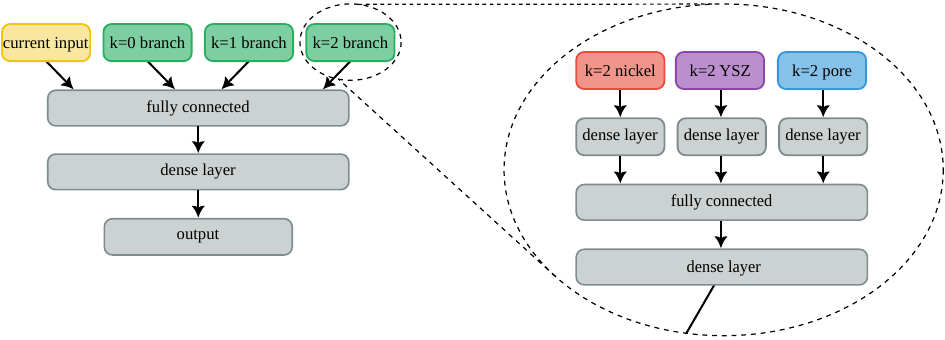}
    \caption{\textbf{Achitecture of ANN used in this research.} ANN used in this research consists of $3$ branches and current density $J$ input. 
    Each branch consists of three inputs corresponding to the $k$th Persistence Diagram $PD_k$ for all of the SOFC phases: nickel, YSZ and pores. Given then, ANN have $10$ inputs}
    \label{fig:network_architecture}
\end{figure*}

The neural network architecture used in this research consists of three main branches corresponding to the $k$-th Persistent Diagram (see \autoref{eq:persistence_diagram}) and input for current density $J$ value, then all of these branches are fully connected to a dense layer with $500$ neurons and then to output layer with $1$ neuron, architecture was presented on \autoref{fig:network_architecture}.
Although we used samples for predefined current values, our model is a single output to predict whole IV characteristics, not only in restricted points. 
Each branch consists of three inputs corresponding to the extracted phases, after each of the inputs there are dense layers fully connected to the other dense layer, the number of neurons in hidden layers is the same for each branch and is $500$ and $1100$ respectively.
Given that, for the $PD_k$ branch, inputs correspond to $PD_k$ for each phase separately.

We proposed such an architecture because we assumed that each branch would better extract the topological features of the microstructure by aggregating PI of the same degree from all phases than by analyzing three Persistence Images of the entire microstructure. Moreover, this learning method allows the model to initially learn the features associated with each phase for a given $k$, subsequently, whole topological information for an assigned $k$ (see \autoref{eq:persistence_diagram}) is aggregated in each branch and finally, 
topological information from all branches and the current value are used to predict microstructure performance.

All models used for performance comparison were trained for $60$ epochs using a mean squared error (MSE) loss function for predicted performance value for each microstructure in training data.
The ADAM\cite{kingma2017adam} optimizer combined with 1cycle scheduling\cite{smith2018disciplined} were used for network parameters optimization, with learning late minimum and maximum respectively \num{5.875e-5} and \num{5.875e-4}.
In order to prevent exploding or vanishing gradients we deploy LeCun kernel initialization.
To avoid overfitting our models, we employ dropout after each dense layer except the one before output, with a dropout ratio of $0.01$.

\begin{table}[ht]
\caption{\textbf{ANN tuned hyperparameters with searched range of values}}\label{tab:hyperparameters}
\begin{tabularx}{\columnwidth}{|lX|}
\hline
\rowcolor[HTML]{9B9B9B} 
Hyperparemeter                        & Value                       \\ \hline
\rowcolor[HTML]{EFEFEF} 
activation function                        & ReLU, SELU, GELU                        \\
\rowcolor[HTML]{FFFFFF} 
first dense neurons                        & one of the value from set $\{200, 350, 500, 1000, 1500, 2000\}$                        \\
\rowcolor[HTML]{EFEFEF} 
second dense neurons                        & value from range $[200, 1500]$ with default value $1000$                        \\
\rowcolor[HTML]{FFFFFF} 
third dense neurons                        & value from range $[200, 1500]$ with default value $1000$                        \\
\rowcolor[HTML]{EFEFEF} 
kernel initializer                         & "he\_normal", "he\_uniform", "lecun\_normal", "lecun\_uniform"                  \\
\rowcolor[HTML]{FFFFFF} 
learning rate                              & value from range $[3\mathrm{e}{-4}, 3\mathrm{e}{-2}]$ with logarithmic sampling \\
\rowcolor[HTML]{EFEFEF} 
optimzer                                   & ADAM, NADAM                                                                     \\
\rowcolor[HTML]{FFFFFF} 
dropout ratio                              & value from range $[0.0, 0.1]$                                                   \\ \hline
\end{tabularx}%
\end{table}

The final hyperparameter values were determined during grid search, \autoref{tab:hyperparameters} presents searched subspace of hyperparameters.

Models were built and trained using Python library \href{https://www.tensorflow.org/}{Tensorflow} version 1.16, and we conducted a hyperparameters search using \href{https://keras.io/keras_tuner/}{KerasTuner}.

\section{Results}
To investigate the potential of combining machine learning and topological data analysis, especially Persistence Homology, we incorporate methodologies from both domains.
We employed deep neural networks and utilized Persistence Images (PI) as topological descriptors of microstructures and then employed a comprehensive set of experiments.

The process of creating a Persistence Image is inextricably linked to the selection parameters of the weight function $w$ and resolution, in this research, we used $w$ as given by \autoref{eq:weight_function}, which have two parameters $C$ and $p$.
Appropriate parameters of weight function depend on the type of data and task, as well as on the amount of noise the data contains.
However, searching appropriate parameters for weight function cleans the data from the noise that may result from the nature of the measuring equipment or method, such as artefacts on MRI caused by interferences, and filters irrelevant topological features, considered as the ones with the short persistence.
In our research we used artificially generated solid oxide fuel cells, therefore data is not affected by measurement noise.

The experimental phase involved a detailed analysis of the influence of PI parameters such as weight function parameters ($C$, $p$) and PI resolution ($m$) on the ANN performance.
After that, we present the generalization potential gained from high-resolution Persistence Images, which were compared with artificial neural networks trained on topological descriptors with a smaller number of pixels.
In the end, we present in detail the results achieved by our representative models clearly demonstrating the advantages of our novel approach and revealing the hidden potential of the usage of machine learning and PH in material science.

\subsection{Weight function parameters}

\begin{figure*}[htbp]
  \centering
  \begin{subfigure}[b]{0.45\textwidth}
    \centering
    \includegraphics{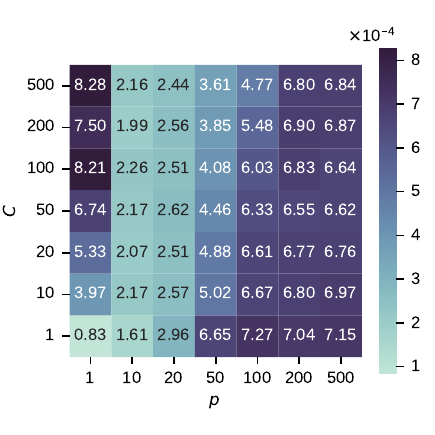}
    \caption{\textbf{Heatmap that illustrates the mean squared error achieved by the model trained on PI with respective parameters of weight function.}
    The X-axis and y-axis represent values of parameters respectively $p$ and $C$.
    The spectrum of the presented results ranges from \num{0.83e-04} to \num{8.28e-04}, where the smallest mean squared error is gained for $p=1$, $C=1$, the largest error is obtained for $p=1$ and $C=500$}
    \label{fig:heatmap_a}
  \end{subfigure}\hfill
  \begin{subfigure}[b]{0.5\textwidth}
    \centering
    \includegraphics{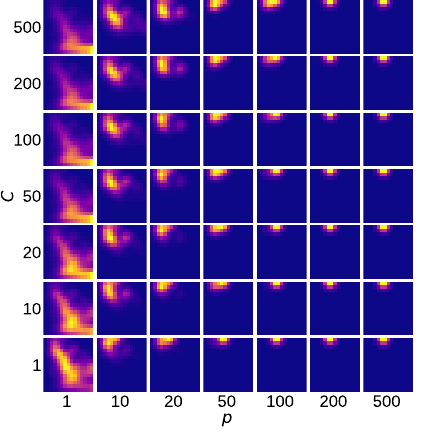}
    \caption{\textbf{Comparison of $k = 0$ Persistence Images for a selected microstructure nickel phase, obtained for different weight function parameters, for the tested values.} Respective PI is presented as a $15 \times 15$ matrix with appropriate weight function parameters $C$ ad $p$. For readability the scale of the color map was normalized for each of the presented descriptors, the light yellow color marks the highest values, while the navy marks values near zero}
    \label{fig:heatmap_b}
  \end{subfigure}
  \caption{\textbf{Model performance and generated PI based on weight function parameters $C$ and $p$}}
  \label{fig:heatmap}
\end{figure*}

In this experiment, we resolved to explore the influence of PI parameters and the associated performance of the artificial neural network.
For this purpose, we trained five neural networks for each combination of parameters $C$ and $p$ from the set $\{1, 10, 20, 50, 100, 200, 500\}$ and took the median of the results for comparison.
The details of neural network architecture and training process are presented in \autoref{sec:ml_details}, for the information about the dataset see \autoref{sec:dataset}.

Squares on \autoref{fig:heatmap_a} correspond to the performance of a specific model measured by mean squared error, whilst \autoref{fig:heatmap_b} shows how the PI changes for the same data when changing the $C$ and $p$ parameters.
The axes for both figures represent the same parameter changes, where the x-axis and y-axis represent values of parameters $p$ and $C$.
The largest and the smallest errors obtained are \num{0.83e-4} and \num{8.28e-4}, respectively, which gives nearly a tenfold difference between these results.
The general trend of model performance is that with an increase of weight function parameters, the model error increases, especially rapid growth noticeable with an increasing value of $p$.
At the same time, it should be noted that for $p=1$ error also grows rapidly with an increasing value of $C$, the observable valley between the results for $p=1$ and $p=20$ is due to the fact that the optimal value should be picked from that space.

\autoref{fig:heatmap_b} illustrates changes that occur in created $k=0$ Persistence Images for a selected microstructure nickel phase, that was chosen randomly from the training set, it can be observed that for $p \geq 50$ all PI are very similar, which results from the nature of the weight function used in this research given by the \autoref{eq:weight_function}, which affects on the intensity of filtration, which zeros out most points while simultaneously enhancing the values of points with sustained persistence.
The high value of both parameters causes more intense filtering, which could eliminate more noise in data and potentially useful information.
The best performance for relatively small values of $C$ and $p$ parameters is due to that the used dataset consists only of artificially generated microstructures which are deprived of significant measurement noise, also, for the high value of parameters, too much data is outfiltered.
Therefore, noticeable stagnation in Persistence Image changes and deterioration in the results for $p \geq 50$ indicate that too much information is filtered from the data.
The parameters values $p=1$ and $C=3$ were used in further analysis.

\subsection{Persistence Image resolution}
\begin{figure}
    \centering
    \includegraphics{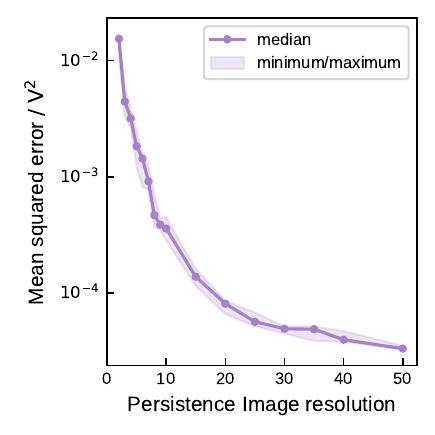}
    \caption{\textbf{ANN mean squared error based on PI resolution.} ANN mean squared error function at PI resolution.
    The points on the figure represent the median value of mean squared error values due to $m$ values from \autoref{eq:vectorized_image}, for which we conducted research and with the light colour range of obtained values was marked. Results obtained for $m < 6$ cannot be considered satisfactory, for $m = 7$ the mean squared error starts to have an acceptable value \num{9.1235e-4}. Best median performance was obtained for $m=50$ with value \num{3.3156e-05}}
    \label{fig:pi_resolution}
\end{figure}
Another parameter having a significant influence to ANN performance is the number of PI pixels.
Since the ANN performance can vary depending on a weights initialization, to provide more balanced results we trained and measured five randomly initialized ANNs for PI resolution $m \in \{$2, 3, 4, 5, 6, 7, 8, 9, 10, 15, 20, 25, 30 , 35, 40, 50$\}$.

Measured performance is shown on the \autoref{fig:pi_resolution} with the mean squared error function on PI resolution. The horizontal axis represents the resolution of Persistence Images, the value of $m$ in \autoref{eq:vectorized_image}, whilst the vertical axis in logarithmic scale, represents model performance measured by a mean squared error.
The solid line represents the median model performance, while the ranges of all measured values are marked in a light colour.
Results obtained for the smallest tested PI resolutions, $m < 6$, cannot be considered satisfactory, due to the fact that mean square error is reduced by 3 orders of magnitude up to the resolution value $m=20$. Above this value, a less significant error reduction can be observed, for $m=7$ the mean squared error starts to have an acceptable value \num{9.1235e-4}.
The best performance, \num{3.3156e-05}, was obtained for $m=50$ leading to a PI size of $2500$.
The good performance of PI for small resolutions ($10 \leq m \leq 20$) is because such a descriptor reduces blending significant features.

Starting from a value of $m=25$ the result did not improve significantly, on average by $\SI{15.8}{\%}$ with each $m$ change, which is substantially less compared to $\SI{36.9}{\%}$ in the range \numrange{2}{20},
whilst at the same time the number of PI pixels starts an increase in respective scopes by $\SI{44.6}{\%}$ and $\SI{56.9}{\%}$, for instance, when $m=50$, the model's achieves a median performance of \num{3.4155e-05}, while the corresponding number of one PI pixels is $2500$, which, including current density value, leads to $22501$ total features for one sample (see \autoref{sec:dataset}). 
Achieving better model performance as the PI resolution increases can be seen as a computability-performance trade-off because as the PI resolution doubles, the number of operations needed to generate them increases fourfold.
Furthermore, the number of model parameters and time required to train the model also increase quadratically with the PI resolution, because the shape of input layers must be adjusted to the PI shape.
In the method presented in this research most time-consuming is data preprocessing, since for each microstructure there is a need to generate nine Persistence Diagrams which is the most time-consuming step in the process of generating Persistence Images.

\subsection{Histograms}
\begin{figure}
    \centering
    \includegraphics{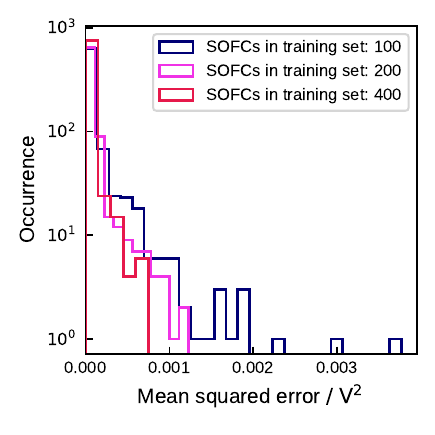}
    \caption{\textbf{Error distribution of models trained on different sizes of the training dataset.} The distribution of mean squared error of the predicted overpotential characteristics by the three models trained on PI with resolution $50$ with different numbers of microstructures}
    \label{fig:histograms}
\end{figure}

Assessing model performance in terms of the size of the dataset, specifically, if similar outcomes can be obtained with less data, was performed by training ANNs on a limited number of microstructures.
Naturally to the machine learning models, increasing the number of samples that the model processes during the training phase, leads to better generalization ability by the model, but in our research, we show that the Artificial Neural Networks trained on the high-resolution PI with small training sets yield comparable results to those trained on the low-resolution PI with extensive training sets.

Results obtained by training ANNs with $100, 200$ and $400$ microstructures in the training set are depicted on \autoref{fig:histograms} with three histograms of a mean squared error on the testing set.
The x-axis of a histogram represents mean squared error values for the testing microstructures, divided into bins, whilst the y-axis corresponds to the count of occurrences for each bin.
With an increasing number of SOFC microstructures in the training set, the model error naturally decreases, indicating better performance.
Models trained with $100$ (navy), $200$ (magenta) microstructures achieve mean squared error respectively \num{1.3073e-4} and \num{7.5835e-05} which are close to the ones obtained with PI resolution respectively $15$ and $20$, the difference between them is $\SI{8.3}{\%}$ and $\SI{23.9}{\%}$.
The percentage error on test data set in the range of \SI{0.98}{\percent} to \SI{20.41}{\percent} when taught on 400 microstructures and between \SI{1.43}{\percent} and \SI{51.59}{\percent} using 100 training samples.

Moreover, it is worth noting the fact that the error of ANN trained on $100$ (navy) microstructures achieves an error comparable to the model trained on $15\times15$ PI but on the $400$ data samples

\subsection{Representative model}

We will now look closer at the results achieved by our representative model, which was trained using Persistence Images with parameters $m=50$, $C=3$, $p=1$.
Our analysis revealed that this model achieved a mean squared error, mean absolute error and $R^2$ score on the test dataset respectively \num{3.3156e-5}, \num{4.0156e-3} and \num{0.9831} reflecting its strong performance.
The mean squared error made by this model is in the range of \num{7.0402e-12} to \num{7.6913e-4}, and the distribution is long-tailed, only for $\SI{7.8}{\percent}$ points mean squared error is greater than \num{1e-4}.

\begin{figure}
    \centering
    \includegraphics{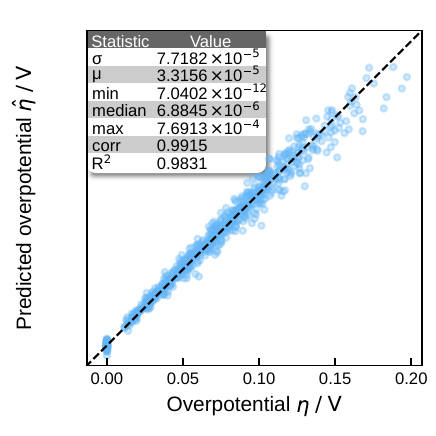}
    \caption{\textbf{Representative model produces high accuracy predictions.} Scatter plot of point $(\eta, \hat{\eta})$, where $\eta$ stands for overpotential computer by numerical mode, whilst $\hat{\eta}$ are overpotential value predicted using ANN}
    \label{fig:scatter}
\end{figure}

The reliability of the ANN trained using homology descriptors is illustrated on the \autoref{fig:scatter} using a scatter plot, where most of the samples are situated on the diagonal demonstrating consistency between the predictions and the ground truth values.
The horizontal axis represents the overpotential $\eta$ given by the numerical model for microstructures separated for tests, whilst the vertical axis corresponds to predicted overpotential $\hat{\eta}$ by the artificial neural network.
Blue markers display the consistency between model outcomes $(\eta, \hat{\eta})$, whereas the dashed line corresponds to the diagonal.
Basic statistics describing the model's error and performance are located in the table in the upper left corner.
The overpotential obtained from numerical model $\eta$, established as truth, is in a strong linear relationship with overpotential predicted with our topological model $\hat{\eta}$ achieving Pearson correlation coefficient equal to \num{0.9915}, indicating high predictive accuracy.
The mean squared errors obtained by the representative model span between \num{7.0402e-12} and \num{7.6913e-4} with median \num{6.8845e-6}, whilst $\sigma = \num{7.7182e-5}$, $\mu = \num{3.3156e-5}$.

Although it is particularly worth looking at errors that the model made for points nearly $\eta =$ \SI{0.0}{\volt} there exist a bunch of samples arranged vertically suggesting systematic error.
Indeed, according to theory, the overpotential of the solid oxide fuel cells for current density $J=\SI{0.0}{\ampere\per\square\metre}$ should then be \SI{0.0}{\volt}, which we took into account in the dataset used in this work, however, the model reaches a value close to zero $(\epsilon = \num{1e-4})$ only for $9$ microstructures out of $100$ making up the test set.
Furthermore, for $J=\SI{0.0}{\ampere\per\square\metre}$ model gives a negative value for $\SI{49}{\%}$ of the test set.
Despite the appearance of these deviations, we do not consider them as significant disadvantages due to that the overpotential for $J=\SI{0.0}{\ampere\per\square\metre}$ has a well-known value equal to $\SI{0.0}{\volt}$.
The value of the Pearson correlation coefficient of all points is \num{0.9915}, this high correlation suggests that the ANN predictions are consistent and reliable with ground truth values. 

Owing to the fact that the scale of overpotential changes for different current densities, we decided to present the basic statistics of error distributions made by the model in each of $J$ value using relative error defined as:
\begin{equation}\label{eq:rre}
    e_i = \frac{\hat{y}_i - y_i}{y_i} 
\end{equation}
for $y_i \neq 0.0$.
Due to the fact that for $J=\SI{0.0}{\ampere\per\square\metre}$ overpotential $\eta$ is \SI{0.0}{\volt}, statistics in these points are presented for residuals.
\autoref{eq:rre} allows us to determine the trend of model estimation.
\autoref{tab:distribution_stats} presents in detail relative error distribution statistics of our representative model achieved on testing microstructures, such as mean and standard derivation, reached by the representative model on the test set for current density $J$ values used in this research.
The last row contains the Pearson (PCC) and Spearman (SCC) correlation coefficients between true and predicted distributions, for $J=\SI{0.0}{\ampere\per\square\metre}$, the value of both coefficients cannot be determined due to the fact that all true values are zero.
Given that for all $J$ values, their means $\mu$ and medians are nearly $0.0$, in general, the model does not have a trend to highly overestimate or underestimate overpotential.

\begin{table}
\caption{\textbf{Basic statistics for relative error reached by the representative model on the test set for current density $J$ values used in this research.} The last two rows contain Pearson (PCC) and Spearman (SCC) correlation coefficients between true and predicted values}
\begin{adjustwidth}{-2cm}{-2cm}
\centering
\label{tab:distribution_stats}
\begin{tabular}{|crrrrrrrr|}
\hline
\rowcolor[HTML]{9B9B9B} 
{\color[HTML]{000000} J/\si{\ampere\per\square\metre}} &
  \multicolumn{1}{c}{\cellcolor[HTML]{9B9B9B}{\color[HTML]{000000} 0}} &
  \multicolumn{1}{c}{\cellcolor[HTML]{9B9B9B}500} &
  \multicolumn{1}{c}{\cellcolor[HTML]{9B9B9B}1000} &
  \multicolumn{1}{c}{\cellcolor[HTML]{9B9B9B}1500} &
  \multicolumn{1}{c}{\cellcolor[HTML]{9B9B9B}2000} &
  \multicolumn{1}{c}{\cellcolor[HTML]{9B9B9B}2500} &
  \multicolumn{1}{c}{\cellcolor[HTML]{9B9B9B}3000} &
  \multicolumn{1}{c|}{\cellcolor[HTML]{9B9B9B}3500} \\ \hline
\rowcolor[HTML]{FFFFFF} \rowcolor[HTML]{FFFFFF} 
$\mu$ & 8.273e-6 & -0.034337 & -0.019676 & -0.015022 & -0.010445 & -0.007758 & -0.008166 & -0.011591  \\
\rowcolor[HTML]{EFEFEF} 
$\sigma$ & 1.879e-03 & 0.113247 & 0.087512 & 0.077548 & 0.071103 & 0.065441 & 0.061552 & 0.060372  \\
\rowcolor[HTML]{FFFFFF} 
min & -4.336e-03 & -0.380086 & -0.319718 & -0.266098 & -0.209135 & -0.175951 & -0.158871 & -0.153399  \\
\rowcolor[HTML]{EFEFEF} 
25th & -1.358e-03 & -0.094742 & -0.067598 & -0.062304 & -0.052102 & -0.047207 & -0.048663 & -0.059586  \\
\rowcolor[HTML]{FFFFFF} 
50th & -1.623e-05 & -0.032957 & -0.017035 & -0.015448 & -0.010772 & -0.007983 & -0.008252 & -0.011782  \\
\rowcolor[HTML]{EFEFEF} 
75th & 1.325e-03 & 0.029198 & 0.035884 & 0.035442 & 0.033922 & 0.032210 & 0.028910 & 0.026976  \\
\rowcolor[HTML]{FFFFFF} 
95th & 3.197e-03 & 0.135981 & 0.122431 & 0.104240 & 0.100446 & 0.092012 & 0.091047 & 0.101246  \\
\rowcolor[HTML]{EFEFEF} 
max & 6.754e-03 & 0.312436 & 0.224737 & 0.195216 & 0.183366 & 0.162252 & 0.143628 & 0.128928  \\
\rowcolor[HTML]{FFFFFF} 
PCC & \multicolumn{1}{c}{\cellcolor[HTML]{FFFFFF}\textbf{--}} & 0.957884 & 0.966457 & 0.966923 & 0.966330 & 0.966595 & 0.966718 & 0.964593  \\
\rowcolor[HTML]{EFEFEF} 
SCC & \multicolumn{1}{c}{\cellcolor[HTML]{EFEFEF}\textbf{--}} & 0.969025 & 0.973789 & 0.974305 & 0.975806 & 0.975038 & 0.975962 & 0.974917  \\
\hline
\end{tabular}%
\end{adjustwidth}
\end{table}

\begin{figure}[htbp]
    \begin{adjustwidth}{-2cm}{-2cm}
        \centering
        \includegraphics{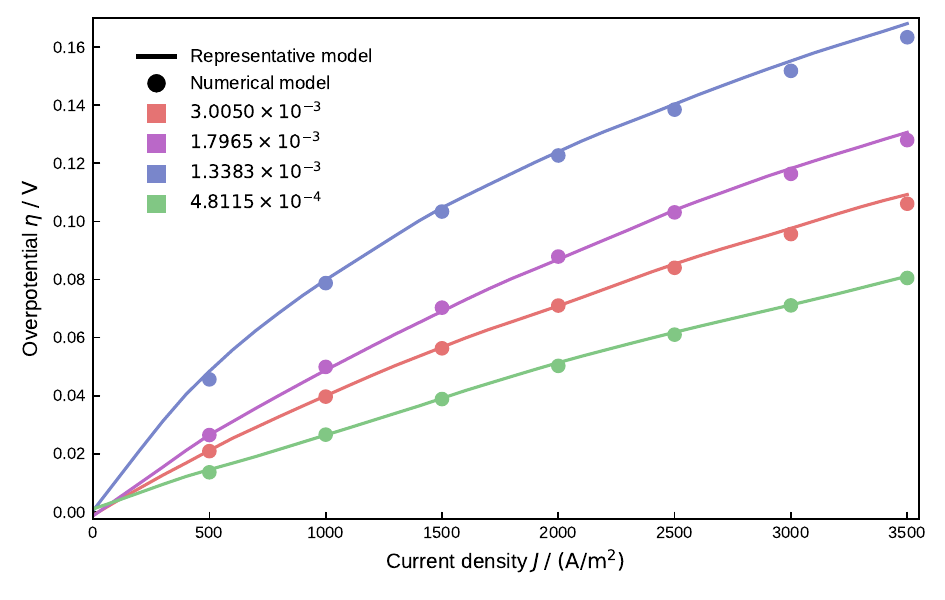}
    \end{adjustwidth}
    \caption{\textbf{Performance curves for four microstructures selected from the test set.} Four curves that illustrate the overpotential characteristics forecasted by representative ANN (solid line) and by a numerical model (dots). Presented microstructures were arbitrally selected from $\SI{50}{\%}$ of samples from the test set for which ANN yielded the lowest error. Root mean squared error (RMSE) for these microstructures is \num{3.0050e-3} (red), \num{1.7965e-3} (purple), \num{1.3383e-3} (blue) and \num{4.6881e-4} (green)}
    \label{fig:rmse_curves}
\end{figure}

\begin{figure}
    \centering
    \includegraphics{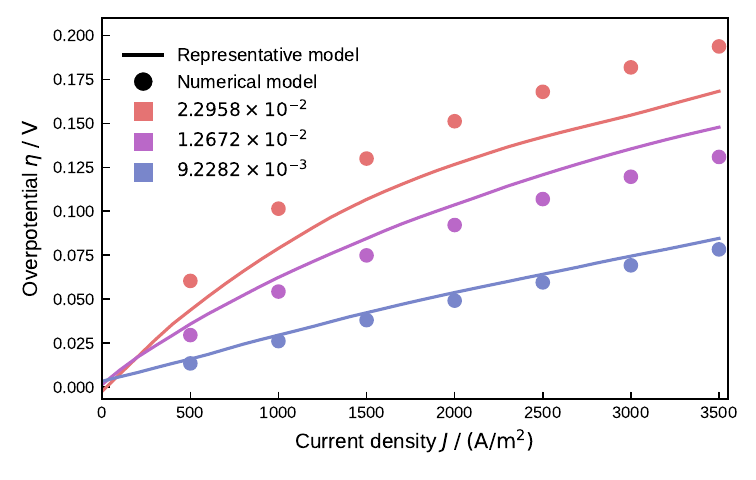}
    \caption{\textbf{Poorly fitted overpotential curves.} Comparison of predicted overpotential (solid line) with numerical model values (dots) for three microstructures from the test set with respectively 1st (red), 5th (purple) and 9th (blue) biggest root mean squared error}
    \label{fig:rmse_worst}
\end{figure}

\autoref{fig:rmse_curves} depicts four overpotential characteristics forecasted by our representative ANN and compared with those by the numerical model.
The horizontal axis represents current density $J$ in the range used in this research, from \num{0} to \SI{3500}{\ampere\per\square\metre} with marked J values used in the training phase.
The vertical axis corresponds to the model's forecasts, where solid lines represent Artificial Neural Network responses, whilst the overpotential achieved through the numerical model is marked with corresponding points.
Value of predicted overpotantial $\hat{\eta}$ spans the range from \num{-2.3194e-3} to \num{0.1634} \si{\volt}.
Presented microstructures were arbitrally selected from $\SI{50}{\%}$ of samples from the test set for which ANN yielded the lowest error.
Characteristics were determined for current density values within the given interval with the step of $\SI{100}{\ampere\per\square\metre}$ and are marked with solid lines to show the overall trend of the curves.
Root mean squared error (RMSE) for these microstructures is \SI{3.0050e-3}{\volt} (red), \SI{1.7965e-3}{\volt} (purple), \SI{1.3383e-3}{\volt} (blue) and \SI{4.6881e-4}{\volt} (green).
The green line shows the microstructure from the test set with the lowest overpotential characteristic RMSE.
For the red characteristic, the phenomenon we mentioned earlier occurs, and predicted values for $J=0$ are negative, hence the range of showed overpotential starts from \SI{-2.3194}{\volt}.

Furthermore, overpotential characteristics RMSE for the whole test set shows promising results because the biggest error is \SI{2.2958e-2}{\volt}, also the $\sigma = \SI{3.6573e-3}{{\volt}}$ and $\mu = \SI{4.4624e-3}{\volt}$. There are only $7$ test samples for which the RMSE is higher than \SI{1e-2}{\volt}.
\autoref{fig:rmse_worst} shows a similar comparison between overpotential values such as \autoref{fig:rmse_curves} using the microstructures with respectively 1st (red), 5th (purple) and 9th (blue) biggest root mean squared error which corresponding error values are \SI{2.2958e-2}{\volt}, \SI{1.2672e-2}{\volt} and \SI{9.2282e-3}{\volt}. The biggest absolute error for a single point between numerical model values and those predicted by ANN occurs at \SI{2500}{\ampere\per\square\metre} and is equivalent to \SI{2.7733e-2}{\volt}, for the microstructure with the biggest RMSE overall. The character of the predicted overpotentials is in agreement with the target characteristics in terms of derivatives, with discrepancies corresponding to the predicted values, indicating overall good quality predictions even in the worst-case scenarios.
\section{Discussion}
The paper investigates the integration of machine learning and topological data analysis, specifically using Persistence Homology to forecast solid oxide fuel cell performance considering its microstructure. We employed deep neural networks and utilized Persistence Images as topological descriptors of microstructures to predict the so-called electrode polarization curves.

The study explored the influence of persistence image (PI) parameters ($C$ and $p$) on the predictive performance of the artificial neural network. The model error increased with higher values of these parameters, especially with an increasing value of $p$. The optimal performance was achieved with relatively small values of $C$ and $p$, specifically $C = 3$ and $p = 1$, due to the nature of the dataset, which consisted of artificially generated microstructures with minimal noise. The representative model, trained using PI with parameters $m = 50$, $C = 3$, and $p = 1$, achieved a mean squared error (MSE) of \num{3.3156e-3} and an $R^2$ score of 0.9831 on the test dataset. These metrics reflect the model's strong performance in predicting electrode overpotential for a given microstructure. 

The performance curves for four selected microstructures from the test set showed varying degrees of deviation from the numerical model, within the range of high-quality simulation, with an $R^2$ score of 0.9831 and a correlation coefficient \num{0.9915}.  Notably, the network prediction agrees well with the test set when considering the function value, the shape of the curve, and derivatives at the end of the considered interval, indicating the usefulness of the neural network in predicting performance or in helping to design the correct electrode microstructure in the future. Moreover, the study demonstrated that ANNs trained on the high-resolution PI with small training sets yielded comparable results to those trained on the low-resolution PI with extensive training sets. This finding is significant as it suggests that high-resolution topological descriptors can enhance model performance even with limited data. This is crucial for applications where data acquisition is expensive or time-consuming. 

By incorporating PH and PI, the research highlights the potential of topological data analysis in capturing essential features of microstructures.
This approach can provide robust descriptors for machine learning models, improving their ability to generalize and predict material properties and thereby helping to improve electrode design in the future.

The integration of machine learning and topological data analysis is a fast-growing field with a wide range of applications in material science, biology, and other domains where understanding the structure-property relationship is crucial. This paper contributes to this field by demonstrating the effectiveness of using Persistence Images as descriptors for training ANNs to predict quantitative answers of the system based on topological (microstructure) and non-topological data (operational conditions), providing a framework for future studies to build upon.

\section*{Author contributions}
\textbf{M.S.}: Methodology, Software, Validation, Formal analysis, Investigation, Data curation, Writing - original draft, Visualization
\textbf{S.B.}: Methodology, Software (Microstructure generation), Validation, Formal analysis, Investigation, Data curation, Writing - original draft, Writing - review \& editing, Visualization
\textbf{T.P.}: Methodology, Software (Fuel cell electrode model), Data curation, Formal analysis, Writing - review \& editing, Visualization
\textbf{G.B.}: Idea, Supervision, Writing - original draft, Writing - review \& editing, Project administration

\section*{Competing interests}
All authors declare no financial or non-financial competing interests. 

\appendix
\section{Microscale charge transport model}
\label{app:micromodel}
The charge transport equations within the electrode were solved using a set of three-dimensional conservation equations coupled with the Butler-Volmer model for the reaction in proximity to the reaction site (triple phase boundary). 

\begin{figure}[H]
    \centering
    \includegraphics[width=0.65\linewidth]{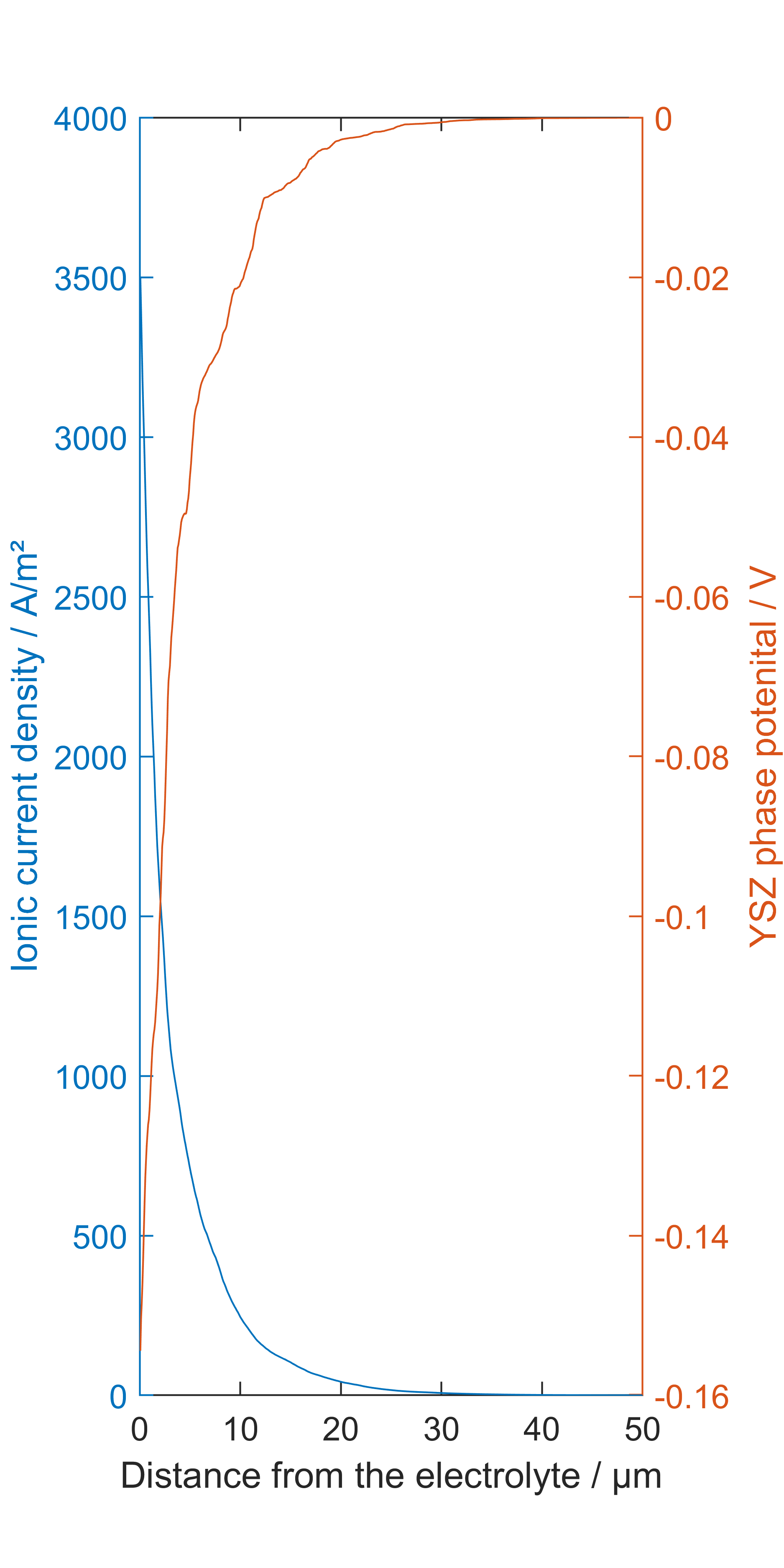}
    \caption{Cross-section averaged ionic current density and potential of the ion-conducting phase in relation to the electron-conducting phase potential $\phi_{\rm ion}$ ($\phi_{\rm el} = 0$). YSZ volume fraction: 30\%, Nickel volume fraction: 30\%, Nickel grain mean radius: $\sim\SI{1}{\micro\metre}$, Current density: $j=\SI{3500}{\ampere \per \metre \squared}$, $T=\SI{1073}{\kelvin}$, $p = \SI{101325}{\pascal}$, $p_{\text{H}_2} : p_{\text{H}_2\text{O}} = 97:3$}
    \label{fig:pot}
\end{figure}

\begin{figure}[H]
    \centering
    \includegraphics[width=0.55\linewidth]{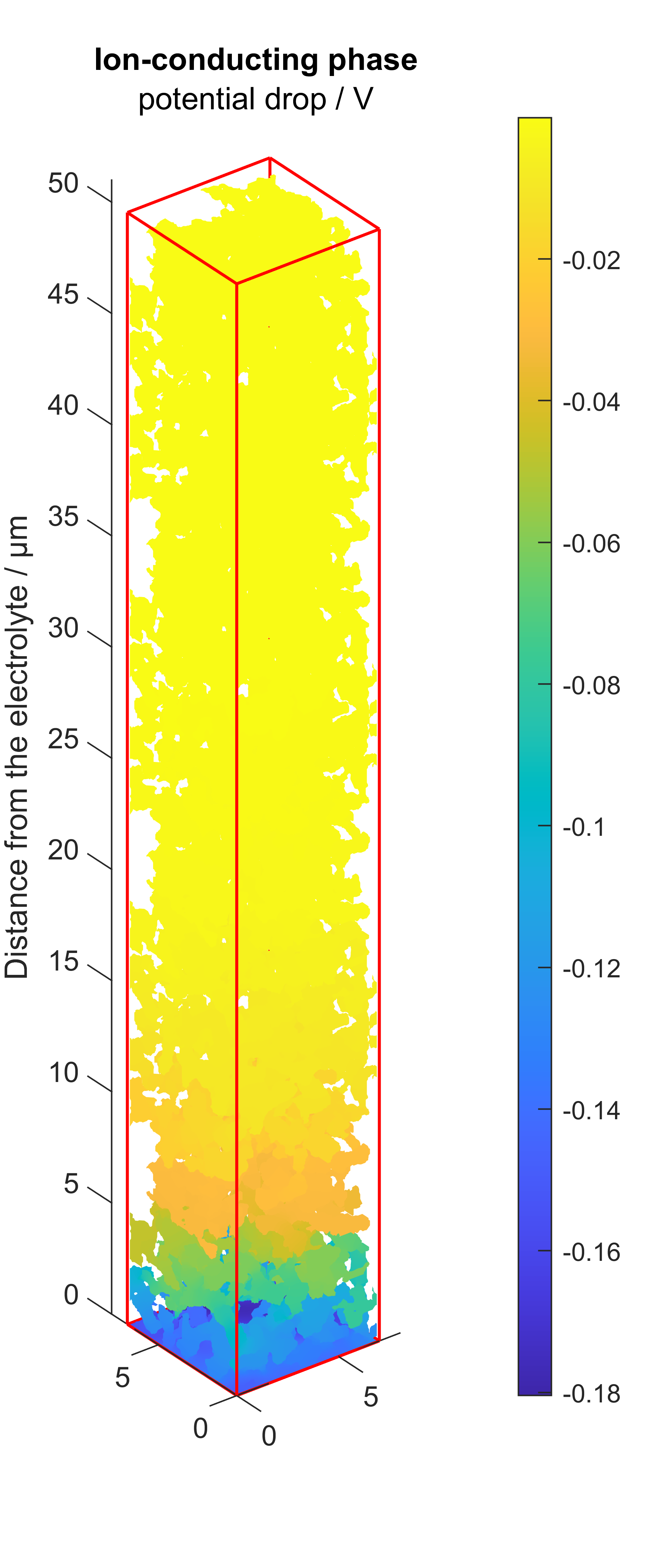}
    \caption{The potential of the ion-conducting phase in relation to the electron-conducting phase potential $\phi_{\rm ion}$ ($\phi_{\rm el} = 0$). YSZ volume fraction: 30\%, Nickel volume fraction: 30\%, Current density: $j=\SI{3500}{\ampere \per \metre \squared}$, Nickel grain mean radius: $\sim\SI{1}{\micro\metre}$ $T=\SI{1073}{\kelvin}$, $p = \SI{101325}{\pascal}$, $p_{\text{H}_2} : p_{\text{H}_2\text{O}} = 97:3$}
    \label{fig:field}
\end{figure}

The model equations can be summed up in the following three dimensional Poisson differential equation:
\begin{equation}
\left\{
\begin{aligned}
\label{eq:poisson3d}
\phi_{\rm{el}} &= \phi_{\rm{el,b}}\\
-i&=\nablavec \dotprod \left( \sigma_{\rm{ion}} \nablavec {\phi_{\rm{ion}}} \right)  \\
i &= i_{0} l_{\rm{tpb}} \dotprod \left [ e^{\frac{\alpha F}{R T} \eta} - e^{-\frac{{\beta} F}{R T} \eta } \right]
\\
\eta &= \phi_{\rm{el}} - \phi_{\rm{ion}}
\end{aligned}
\right.
\end{equation}
where $\phi$ (V) is the electric potential, $\sigma_{\mathrm{ion}}$ (\SI{}{\siemens \per \metre}), is the ion conductivity $\alpha$ is the  forward charge transfer coefficient, $\beta$ is the backward charge transfer coefficient ($\alpha=2$, $\beta=1$). $F \ (\SI{}{\ampere \second \per \mole})$ is the Faraday's constant. The $l_{\rm tpb} (\SI{}{\metre \per \metre \cubed})$ is the triple phase boundary density. The $i_0 \ (\SI{}{\ampere \per \metre \squared})$ is the equilibrium exchange current density, based on the experimental data by de Boer \cite{deBoer1998, kanno2011simulation}. In an interpolated form,
\begin{equation}
    i_0 = 31.4 \cdot p_{\text{H}_2}^{-0.03} \cdot p_{\text{H}_2\text{O}}^{0.4} \cdot \exp\left(-\frac{18300}{T}\right)
\end{equation}
$p_k$ (Pa) is the partial pressure of reagent $k$. $\sigma_{\mathrm{ion}}$ is computed from the following equation:
\begin{equation}
    \sigma_{\text{ion}} = 3.34 \times 10^4 \cdot \exp\left(-\frac{10300.0}{T}\right)
\end{equation}
using experimental data from literature \cite{anselmitamburini1998, kanno2011simulation}. The computations were carried out for the following conditions: $T=\SI{1073}{\kelvin}$, $p = \SI{101325}{\pascal}$, $p_{\text{H}_2} : p_{\text{H}_2\text{O}} = 97:3$, which are realistic for SOFC cells. The model was applied to three-dimensional digital material representations (DMR) generated within the scope of the present study. The finite difference method (FVM) was used for discretizing the computational domain, in such a way that each voxel of the conducting phase in the DMR corresponded to a single FVM node. It is worth noting that in the present model $l_{\rm tpb}$ is not distributed evenly, but rather contains non-zero values only in nodes neighboring the triple phase boundary. The triple phase boundary density for the DMRs was computed using the volume expansion method \cite{Iwai2010}. The resulting set of of nonlinear equations is solved by means of successive substitution. First, the non-linear component (arising from the generation term in Poisson equation \ref{eq:poisson3d}) is linearized. The resulting linear equation set is solved using the preconditioned conjugate gradient method with incomplete LU factorization preconditioning, as implemented in MATLAB 2023b (The MathWorks, Inc.) \cite{Barrett1994}. Using the output, new linearization coefficients were computed, and the corrected linear equation set was solved again. The iterative process was repeated until convergence. Example results are presented in Figures \ref{fig:pot}, and \ref{fig:field}. It can be seen that the active layer, indicated by sharp gradient of potential and current density concetrates in proximity of the electrolyte. Irregularities in the presented potential field stem from the non-homogeneous distribution of conducting phases and reaction sites withing the modeled material.
\newpage


\section*{Data availability}
The datasets generated and analysed during the current study are available in the \href{https://zenodo.org/records/13731825?preview=1&token=eyJhbGciOiJIUzUxMiJ9.eyJpZCI6IjgwMjQyN2FhLWU4YzAtNDk1NC1hNWE2LTg5MDZkNjdhNDNlMyIsImRhdGEiOnt9LCJyYW5kb20iOiJiZDJjMzUzODVhZDk4YjY2MDBjZjVjZDc1MjY2MjhkMiJ9.vEgZumvoNH0xjqzOqMXGlsusDVk6ykHAnw8Va9lksPHJJUHa6CDh3v-fryuZZohYoO5Ku8AWiXcGfmn13MJ7xg}{Zenodo repository}. 

\section*{Code availability}
All Python code for this study is available on \href{https://github.com/MaksymSz/Computational-Materials-publication}{github repository}.

\section*{Acknowledgements}
The authors acknowledge support by the program ``Excellence Initiative---Research University'' for the AGH University of Krakow, Poland.
We gratefully acknowledge Polish high-performance computing infrastructure PLGrid (HPC Center: ACK Cyfronet AGH) for providing computer facilities and support within computational grant no. PLG/2024/017072

\bibliographystyle{elsarticle-num}
\bibliography{bibliography}

\end{document}